# High pulse energy supercontinuum laser for spectroscopic photoacoustic imaging of lipids in the 1650-1850 nm window

**MANOJ KUMAR DASA,**[1,*] **CHRISTOS MARKOS,**[1] **CHRISTIAN. R. PETERSEN,**[1] **AND OLE BANG**[1,2]

[1]*DTU Fotonik, Department of Photonics Engineering, Technical University of Denmark, 2800 Kgs. Lyngby, Denmark*
[2]*NKT Photonics A/S, Blokken 84, Birkerød 3460, Denmark*
*\*manda@fotonik.dtu.dk*

**Abstract:** Detection and identification of lipids are highly coveted for the interrogation of chronic diseases such as atherosclerosis and myocardial infarction. Intravascular photoacoustic imaging (IVPA) and deep tissue imaging are modern techniques, which rely on complex near infrared (NIR) optical parametric oscillators (OPOs) and other high-power solid-state laser systems for the diagnosis, which in turn make the systems bulky and expensive. In this work, we propose a cost-effective directly modulated high pulse energy supercontinuum source (operating in kHz regime) based on a standard optical fiber with pulse energy density as high as ~ 26 nJ/nm. We demonstrate how such supercontinuum source combined with a tunable filter can be highly suitable for vibration-based photoacoustic imaging and spectroscopy of lipids in the molecular overtone band of lipids (1650-1850 nm). We show the successful discrimination of two different lipids (cholesterol and lipid in adipose tissue) and the photoacoustic cross-sectional scan of lipid-rich adipose tissue at three different locations. The proposed high pulse energy supercontinuum laser paves a new direction towards compact, broadband and cost-effective source for multi-spectral spectroscopic photoacoustic imaging.

## 1. Introduction

Photoacoustic imaging (PAI) is a hybrid imaging modality offering label-free high contrast imaging combined with high spatial resolution based on the wavelength-dependent molecular absorption and ultrasonic detection, respectively [1]. The ability of PAI to discriminate various endogenous agents using absorption contrast makes it a promising technique for the detection, diagnosis, and monitoring of various diseases [2-4]. Lipids are such key endogenous agents inside the body, as they act as major contrast agents in the identification of fatal chronic diseases like atherosclerosis and myocardial infarction [5]. The wavelengths used for PAI in most of the aforementioned cases fall in the range of UV-VIS part of the electromagnetic spectrum, due to the prominent absorption of the major endogenous agents (originating within the biological tissue) like hemoglobin, melanin, etc. [3]. Therefore, most of the early studies have been focused on photoacoustic detection of lipids in the 420-530 nm [5] and 680-900 nm wavelength range [6]. However, operation in the UV and VIS part of the spectrum provides a high spatial resolution but limited penetration depth, due to strong absorption of hemoglobin inside the blood. Thereby, enabling the need for saline flush during the imaging. One possible way to achieve higher penetration depths is to operate at longer wavelengths, especially in the overtone bands of C-H vibration bonds in lipids (1100-1300 and 1650-1850 nm) [7-11] as the absorption spectrum of lipid show well-differentiated peaks compared to the other constituents present inside the biological tissue. Moreover, the absorption of hemoglobin in these bands is relatively low, thereby eliminating the need for saline flush during the imaging of the lipids [11].

Photoacoustic detection and identification of lipids in the overtone bands (1100-1300 and 1650-1850 nm) are currently performed using optical parametric oscillators (OPO's) and yttrium aluminum garnet (YAG) lasers. However, these sources have a high cost, large footprints, low pulse repetition rates and therefore the idea of developing compact and portable PAI systems is basically unachievable. It has been already reported that that the above-mentioned laser sources can be circumvented to a certain extent using broadband sources based on stimulated Raman scattering (SRS), four-wave mixing (FWM) [12-14] and supercontinuum (SC) generation [15-16]. However, these sources fail to either emit light or to deliver enough pulse energies in the required broadband spectral region, so as to make them suitable for photoacoustic detection and identification of lipids in the overtone bands.

In the scope of this report, we propose for the first time a cost-effective high pulse energy SC source for spectroscopic photoacoustic imaging, using a low-cost directly modulated laser diode (at 1550 nm) and few meters of standard single mode optical fiber (SMF). We demonstrate SC with pulse energy densities of about 26 nJ/nm over a broad bandwidth from 1500 up to 1950 nm which at the moment is the record among every commercial source. We also demonstrate the application of the proposed SC source in conjunction with a linear variable filter (LVF) for photoacoustic identification and detection of lipids in the first overtone region (1650-1850 nm).

## 2. Methods

2.1  High-pulse energy supercontinuum laser source

Figure 1 (a) shows the schematic of the home-built SC laser source. The laser system primarily consists of a 1550 nm based fiber coupled directly modulated laser diode (MLT-PL-R-OEM20, MANLIGHT), emitting pulses with duration 3 ns (FWHM) and energy of about 30 µJ at a pulse repetition rate of 30 kHz driven by an external function generator (TG2000, AIM-TTi).

The pulses from the laser diode are then used to pump a standard commercially available SMF. The fiber has a numerical aperture of 0.14 and a zero dispersion wavelength (ZDW) close to 1300 nm, resulting in the anomalous dispersion at the pump wavelength. In order to avoid the back reflections from the end facet of the fiber, a bend filter was implemented and the fiber was angle cleaved (setting a cleaving angle of 8° in the tension cleaver (CT-100, FUJIKURA)). The fiber is connectorized (FC/APC, THORLABS) for the ease of handling.

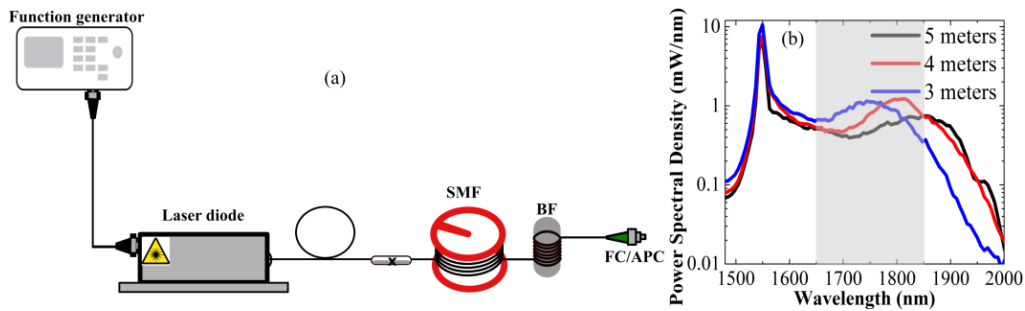

**Figure 1(a).** Schematic of the high-pulse energy SC laser source. An external function generator is used to drive a 1550 nm based fiber coupled directly modulated laser diode, emitting pulses (3 ns) with a repetition rate of 30 kHz. The output from the laser diode is used to pump standard SMF. A bend filter (BF) is implemented and the fiber is angled cleaved, so as to eliminate any back reflections from the end facet. Figure 1(b). SC spectra generated by pumping three different lengths of SMF (5, 4 and 3 meters). The highlighted region showing the first C-H overtone region in lipids (1650-1850 nm).

Three different lengths of SMF's (5, 4 and 3 meters) were tested and the output spectra from the fibers were measured using an array spectrometer (NIRQuest512, OCEAN OPTICS). The output spectra of the SC generated in all the three cases are shown in figure 1(b).

When intense nanosecond laser pulses are pumped into an SMF, the strong confinement and long interaction lengths of the fiber induce nonlinear spectral broadening, thereby yielding a broadband output spectrum referred to as SC. The broadening obtained by pumping the fiber in the anomalous regime is attributed to the initial phase matched four-wave mixing (FWM) and the subsequent effects during the pulse propagation. Particularly, the combination of FWM and Raman scattering perturbs the waveform causing the onset of modulation instability (MI). MI causes the breakup of the pulse into a sea of solitons, which extend the SC spectrum towards longer wavelengths due to the soliton self-frequency shifting [17].

In this work, we are aiming to apply spectroscopic photoacoustic imaging of lipids in the first overtone region of C-H vibration bonds lying in the 1650-1850 nm band. Since the SC is directly related to the propagation length of the fiber, we, therefore, optimized the output of the SC for the spectral region of 1650-1850 nm by reducing the length of the fiber. From figure 2 (a), it can be seen that 3 meters of fiber is enough to generate SC in the required first overtone region of lipids with pulse energy densities significantly higher (within the region of interest) than the commercial SC laser (SuperK COMPACT, NKT Photonics).

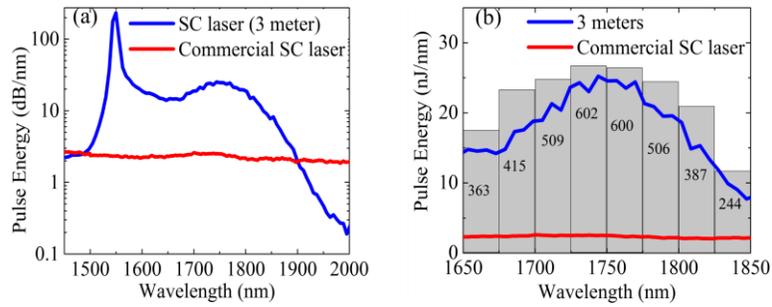

**Figure 2.** Output spectrum of high-pulse energy SC laser source (a) Pulse energy density of home built SC laser (3 meters of SMF) in comparison with commercial SC laser (b) Pulse energy densities of the SC laser in the first overtone region (3 meters of SMF) in comparison with the commercial SC laser, the inset shows the measured pulse energies in 25 nm bands.

Figure 2(b) shows the pulse energy densities of our custom-made SC laser (3 meters of SMF) in direct comparison to the commercial SC laser in the first overtone region of lipids. The inset (vertical bars) in figure 2(b) specify the measured pulse energies in 25 nm bands which are sufficiently high for typical optical resolution based photoacoustic imaging [13, 16].

2.2   Spectroscopic photoacoustic imaging system

The experimental setup of the transmission-based spectroscopic photoacoustic imaging system is shown in figure 3. The system employs the aforementioned high pulse energy SC laser source in conjunction with an LVF as the tunable excitation source. The output from the laser is collimated using an aspheric lens (L1) (A220TM, THORLABS) and then sent to the LVF (Vortex Optical Coatings Ltd) which covers from 1200 nm up to 2500 nm with a bandwidth of 2% of the peak wavelength. The filtered light is then steered using the mirrors (M1, M2, and M4) and focused on the sample using a 5x objective lens (L2). A flip mirror mount, a biconvex lens (L3), and a CCD camera are used to optically image and align the sample using ambient light. The major part of the excitation is built in a cage system, in order to make the setup compact and robust.

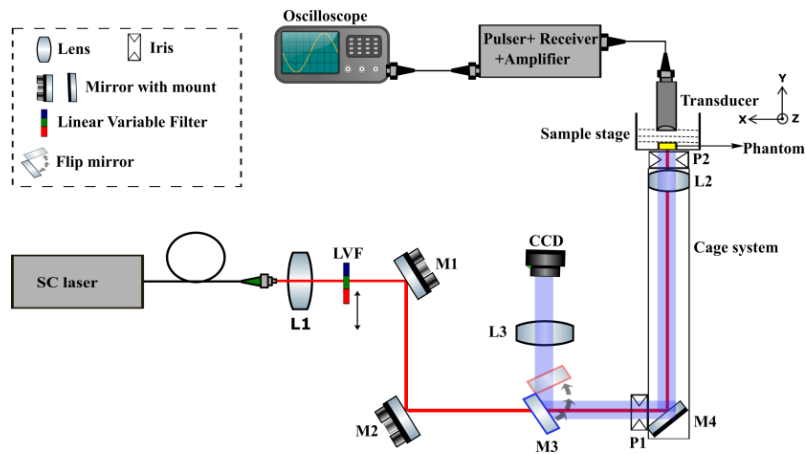

**Figure 3.** Schematic of the spectroscopic photoacoustic imaging system. Fiber-coupled SC laser is used an excitation source, the light from the laser is collimated using an objective lens (L1). The excitation wavelength is filtered using a LVF. The filtered light is steered using the mirrors (M1, M2, and M4) and then focused onto the phantom using an objective lens (L2). The generated photoacoustic signal is detected using a focused transducer, amplified using low-noise amplifiers and then sent to an oscilloscope. P1 and P2 are pinholes. A flip-mirror (M3), a lens (L3) and a CCD are used to align the sample

The generated photoacoustic signal is detected using a commercial 25 MHz ultrasonic transducer (Optel, Poland), with a 3 mm lead zirconate titanate (PZT) crystal as the piezoelectric active element. The acquired signals are filtered by a low pass filter (BLP-50+, MINI-CIRCUITS), amplified using a low-noise wideband signal amplifier (ZFL-500LN, MINI-CIRCUITS) and are then sent to an oscilloscope (HDO9404, TELEDYNE LECROY). Phantoms are placed in a dish, with a 1 mm thick microscope slide fixed at the bottom of the dish, so that the dish is transparent to the optical excitation. The Phantoms are immersed in distilled water, which act as a coupling medium for the generated acoustic waves. An x-y-z high precision stage is used for the alignment of the transducer with the confocal plane of the excitation and detection. An x-stage is used to align and move the sample to the confocal plane.

## 3. Results

3.1 Identification of lipids using photoacoustic spectroscopy

Photoacoustic spectra of two different lipids (commercial grade cholesterol and the lipid in adipose tissue) were measured in 1600-1800 nm wavelength range, with a step size of 20 nm. The excitation bands filtered using LVF, their FWHM and the pulse energies with respect to the center wavelength are shown in figure 4(a) and figure 4(b) respectively. It can be observed from figure 4(b) that the measured pulse energies of the excitation bands (>380 nJ) are sufficiently high to excite photoacoustic signal.

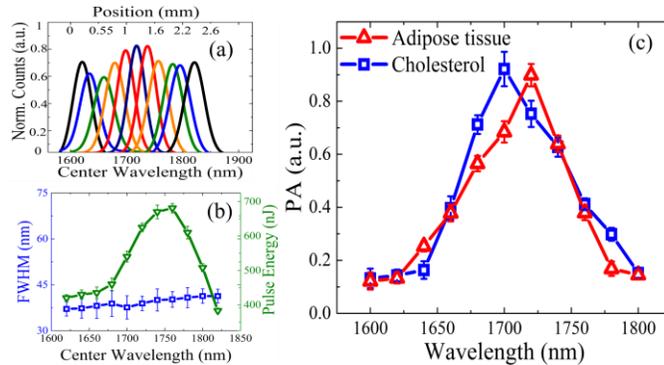

**Figure 4**(a) Excitation bands filtered using LVF. 4(b) FWHM and pulse energies of the excitation bands with respect to the center wavelength. 4(c) Photoacoustic spectra (normalized) of commercial grade cholesterol and adipose tissue in the 1600-1800 nm wavelength range

PA signals detected by the transducer are low-noise amplified (58 dB), averaged over 1000 pulses, and normalized to the reference measurement and also to the pulse energies of the excitation bands. However, the pulse to pulse intensity fluctuations of the SC laser are not considered as we average over 1000 pulses. The corresponding lipid spectra are overlaid in a single plot and shown in figure 4(c). It can be observed that there are dominant differences in the photoacoustic spectra of the lipids, which are attributed to the changes in their respective molecular structures, thereby enabling for the identification of the specific type of lipid. Moreover, the shift in the peak location of the different lipids (1700 nm and 1720 nm) is in good agreement with the earlier findings [10, 17].

3.2 Detection of lipid using photoacoustic imaging

Photoacoustic cross-sectional scans were performed on a 2.5 mm thick lipid-rich adipose tissue obtained from the chicken. 1720 nm center wavelength band is chosen as excitation mainly because the PA signal obtained at 1720 nm band was more prominent when compared to other wavelength bands (refer to figure 4(c)). The measured pulse energy at the sample plane was about 650 nJ.

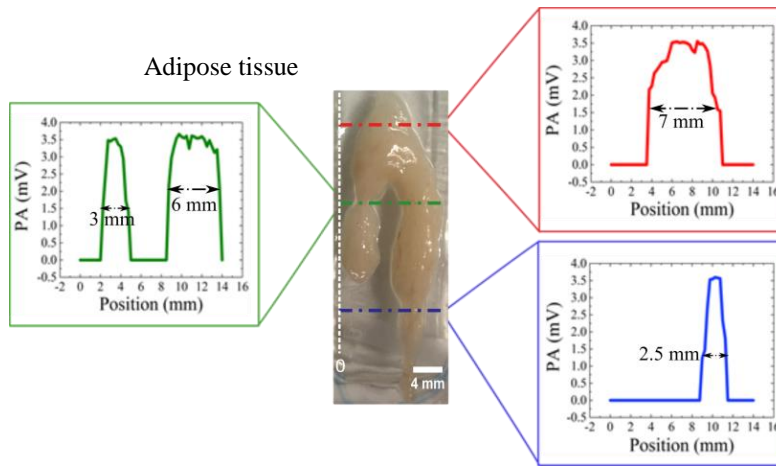

**Figure 5.** Optical image of the lipid-rich adipose tissue from chicken. Insets show photoacoustic cross-sectional scan profiles along the dotted lines red, green and blue respectively

The detected PA signals at every excitation point of the sample were amplified by 58 dB, averaged over 100 pulses and normalized with the reference signal (obtained without the sample). From the insets, it can be clearly observed that the structural information of the tissue can be obtained from the PA cross-sectional scans. Moreover, the PA scans along the dotted lines correspond well with the structure of the tissue sample.

## 4. Conclusion

In summary, we have demonstrated the development of a novel cost-effective high pulse energy SC laser source with a pulse density as high as ~ 26 nJ/nm based on 3 meters of standard SMF and a low-cost telecom range diode laser. We show how the proposed source can be used for the photoacoustic identification and detection of lipids in adipose-rich chicken tissue in the first overtone region (1650-1850 nm). The approach in this report can be readily adapted to scale the output pulse energies to 10's of µJ using higher peak power pump lasers, making them promising sources for the in-vivo SPAI applications.

## 5. Acknowledgements

The research project leading to this work has received funding from the European Union's Horizon 2020 research and innovation programme under the Marie Sklodowska-Curie grant agreement No 722380. C.M. acknowledges support from the Danish Council for Independent Research (4184-00359B).